\input harvmac

\def\half{{1\over 2}}

\def\e{\epsilon}

\def\a {{\alpha}}
\def\l {{\lambda}}

\def\b {{\beta}}
\def\c {{\gamma}}
\def\g {{\gamma}}
\def\d {{\delta}}
\def\s {{\sigma}}

\def\t {{\theta}}
\def\tb {{\bar\theta}}

\def\p {{\partial}}
\def\N {{\nabla}}

\Title{\vbox{\hbox{ IFT-P.059/99}}}
{\vbox{ \centerline{Quantization of the Type II Superstring in a}
\centerline{ Curved Six-Dimensional Background}}}
\centerline{Nathan Berkovits}
\vskip .2in
\centerline{\it Instituto de F\'{\i}sica Te\'orica,
Universidade Estadual Paulista}
\centerline{\it Rua Pamplona 145, 01405-900,
S\~ao Paulo, SP, Brasil}
\vskip .5in

\centerline{\bf Abstract}
A sigma model action with N=2 D=6 superspace variables
is constructed for the Type II superstring
compactified to six curved dimensions with Ramond-Ramond flux. 
The action can be quantized
since the sigma model
is linear when the six-dimensional spacetime is flat. When the 
six-dimensional spacetime is $AdS_3\times S^3$, the action
reduces to one found earlier with Vafa and Witten.

\vskip .2in
\Date{August 1999}

\newsec{Introduction}

Construction of quantizable
superstring actions in $AdS_d\times S^d$
backgrounds with Ramond-Ramond flux
is currently of great interest. Although such a construction
has not yet been done for the $AdS_5\times S^5$ case,
it has been done for the $AdS_2\times
S^2$ and $AdS_3\times S^3$ cases. In the $AdS_2\times S^2$
case \ref\BBZ{N. Berkovits, 
M. Bershadsky,
T. Hauer, S. Zhukov and B. Zwiebach, {\it Superstring
Theory on $AdS_2\times S^2$ as a Supersymmetric Coset},
hep-th/9907200.},
this construction was straightforward since a 
quantizable sigma
model action was already known for the superstring
in a general four-dimensional background \ref\siegel
{N. Berkovits and
W. Siegel, {\it Superspace Effective Actions for 4D Compactifications
of Heterotic and Type II Superstrings}, Nucl. Phys.
B462 (1996) 213, hep-th/9510106.}.
In the $AdS_3\times
S^3$ case \ref\VW 
{N. Berkovits, C. Vafa and  E. Witten,
{\it Conformal Field Theory
of AdS Background with Ramond-Ramond Flux},
JHEP 9903 (1999), hep-th/9902098.},
however, guesswork had to be used since the
action in a general six-dimensional background had not
yet been found.

In this paper, a quantizable 
sigma model action is constructed
for the superstring in a general six-dimensional background.
This construction is useful for several reasons.
Firstly, it allows quantization of the 
superstring in six-dimensional curved backgrounds
other than $AdS_3\times S^3$. Secondly, it provides
a new and simpler description of the 
$AdS_3\times S^3$ action which may be useful for
the construction of vertex operators. Thirdly, it
provides clues which might be useful for quantization 
of the superstring in a general ten-dimensional background.

The six-dimensional action in a flat background
was constructed in
\ref\vafa{N. Berkovits and C. Vafa,
{\it $N=4$ Topological Strings}, Nucl. Phys. B433 (1995) 123,
hep-th/9407190.}(and reviewed in \VW)
using worldsheet variables from the `hybrid' description
of the superstring.
These hybrid variables are related by
a field redefinition to the worldsheet variables
of the Ramond-Neveu-Schwarz (RNS) formalism and include
spacetime spinors as in
the Green-Schwarz (GS) formalism. The hybrid action has
an N=2 worldsheet superconformal invariance which replaces
$\kappa$ symmetry of the GS action and which is
related to a twisted N=2 BRST symmetry
of the RNS formalism \ref\review{
N. Berkovits, {\it A New Description Of The Superstring},
Jorge Swieca Summer School 1995, p. 490, hep-th/9604123.}.
Unlike the GS action, quantization is
straightforward since the hybrid action in a flat
background is quadratic.
 
In the formalism of \vafa\VW, only half of the sixteen $\t$'s
of N=2 D=6 superspace 
are present as fundamental worldsheet fields. Although
this preserves manifest SO(5,1) Lorentz invariance,
it breaks
half of the manifest D=6 supersymmetries.
This fact made it difficult to generalize the formalism
to arbitrary curved six-dimensional backgrounds.

In this paper, this difficulty will be overcome by
adding eight more $\t$'s (and their conjugate momenta)
as fundamental fields in
the action, as well as eight first-class ``harmonic'' constraints
\ref\galp
{A. Galperin, E. Ivanov, S. Kalitsyn, V. Ogievetsky and
E. Sokatchev, {\it Unconstrained N=2 Matter, Yang-Mills and Supergravity
Theories in Harmonic Superspace}, Class. Quant. Grav. 1 (1984) 469.}
\ref\sok{
F. Delduc and E. Sokatchev, {\it 
Superparticle with Extended Worldline Supersymmetry},
Class. Quant. Grav. 9 (1992) 361.} 
which can be used to gauge away these new fields.
With these additional fields and constraints, it
will be easy to construct a sigma model action for
the superstring in an arbitrary curved six-dimensional
background.

Like the GS sigma model action, the hybrid action is
defined using superspace variables which are
extremely convenient for describing 
backgrounds with Ramond-Ramond flux. However,
unlike the GS sigma model action, the hybrid action 
reduces to a free quadratic action when the six-dimensional
spacetime is flat.  By using a normal coordinate
expansion, this allows quantization
in a curved background. Another difference with
the GS action is that the hybrid action
contains a coupling of the spacetime dilaton to the
worldsheet curvature, as expected from the coupling-constant
dependence of scattering amplitudes.

In section 2 of this paper, the hybrid
action will be reviewed in a flat six-dimensional background.
In section 3, this action will be written 
in a manifestly N=2 D=6 supersymmetric form
by introducing new $\t$ variables and new
harmonic constraints \galp\sok.
In section 4, the manifestly spacetime-supersymmetric form of the 
action will be generalized to a curved
six-dimensional background. And in section 5, the
action will be shown to reduce to that of \VW\ when the
six-dimensional background is chosen to be $AdS_3\times
S^3$ with Ramond-Ramond flux. 

\newsec{Review of hybrid action in a flat six-dimensional background}

\subsec{Action and N=2 constraints}

In a flat six-dimensional
background, the hybrid formalism was developed
in reference \vafa\
and was reviewed in \VW. Besides $x^m$ for $m=0$ to 5,
the six-dimensional left-moving worldsheet fields consist of eight fermions,
$\t^\a$ and $p_\a$ for $\a=1$ to 4, and two chiral bosons, $\rho$ and $\s$.
For the closed superstring, the right-moving worldsheet fields consist of
$\tb^{\bar \a}$ and $\bar p_{\bar \a}$ for $\bar\a=1$ to 4, 
and two anti-chiral bosons, $\bar\rho$ and $\bar\s$.
For the Type IIB (or Type IIA) superstring, an up $\a$ index and 
up (or down) $\bar \a$
index transform as 4 representations of SU(4), and a down $\a$ index
and down (or up) $\bar \a$ transform as $\bar 4$ representations.
In addition, one has a $c=6$ N=2 superconformal field theory representing
the compactification manifold.

In a flat background, the free action is 
\eqn\freeaction{S=\int d^2 z (\half
 \p x^m \bar\p x_m + p_\a\bar\p\t^\a + \bar p_{\bar \a}
\p\bar \t^{\bar \a} ) + S_B+S_C}
where $S_C$ is the action for the compactification variables and
$S_B$ is an action for the chiral and anti-chiral
bosons which we will not write explicitly. One also has
the following critical N=2 superconformal generators:
$$T=   
\half\p x^m \p x_m +
p_\a\p \t^\a +\half\p\rho\p\rho
+\half\p\sigma\p\sigma
+{3\over 2}\p^2 (\rho+i\sigma)+T_{C},$$
\eqn\freeconstraints{ G^{+} = 
- e^{-2\rho-i\sigma} (p)^4 ~ +{i\over 2}
e^{-\rho} p_\a p_\b \p x^{\a\b} }
$$
+e^{i\sigma}( \half\p x^m \p x_m +
p_\a\p \t^\a  +\half\p(\rho+i\sigma)\p(\rho+i\sigma)-\half\p^2(\rho+i\sigma))
+ G^{+}_{C} , $$
$$G^{-}  =
e^{-i\sigma}+ G^{-}_C ,$$
$$J=\p(\rho+i\sigma)~+J_{C} ,$$
where 
$[T_C,G^+_C,G^-_C,J_C]$ are the $c=6$ N=2 generators of the
superconformal field theory representing the compactification,
$(p)^4={1\over 24}\e^{\a\b\c\d}p_\a p_\b p_\c p_\d$, 
$x^m$ has been written in bispinor
notation as  
$x^{\a\b}=(\sigma_m)^{\a\b} x^m$, and $(\s_m)^{\a\b}$
are the six-dimensional Pauli matrices satisfying
$$(\sigma_m)^{\a\b}(\sigma_n)_{\b\c} +
(\sigma_n)^{\a\b}(\sigma_m)_{\b\c} =2\eta_{mn}\d^\a_\g$$
with
$(\s_m)_{\a\b}$ defined as 
$(\s_m)_{\a\b} 
=\half\e_{\a\b\c\d} (\s_m)^{\c\d}.$ 
As described in \vafa, these worldsheet variables can
be obtained from the RNS worldsheet variables by
a field redefinition and the constraints of \freeconstraints\
are related to the stress tensor, BRST current, $b$ ghost,
and ghost-number current of the RNS formalism.

\subsec{Massless compactification-independent vertex operators}

Since the integrated 
form of the massless
vertex operators appear in the sigma model
action, it will be useful to review these operators here, beginning with
the simpler case of open string massless vertex operators.

The massless compactification-independent open string 
vertex operators are described by a superfield $V(x,\t,\rho+i\s)$
which can be expanded as 
$V= \sum_n e^{n(\rho+i\s)} V_n(x,\t)$ where $V_n(x,\t)$ is
an arbitrary function of the zero modes
of $x^m$ and $\t^\a$. The chiral bosons only appear in the combination
$\rho+i\s$ in order that $V$ has no poles with $J$.
The integrated form of the vertex operator is given
by
\eqn\intopenver{U= \int dz  
G^- G^+ V(x,\t,\rho+i\s)}
where $G^\pm Y$ will always denote the single pole in the OPE of 
$G^\pm$ with $Y$. Using the fact that $G^- e^{n(\rho+i\s)}=0$
for $n\leq 0$, one can show that when $V$ is on-shell, 
\eqn\udef{U= \int dz 
 [-  {{\e^{\a\b\c\d}}\over 6}
e^{-\rho-i\s} p_a ~(\N_\b\N_\c\N_\d ) +i
  p_\a ~(\N_\b \partial^{\a\b} ) +{i\over 2}
 \p x^{\a\b} ~(\N_\a\N_\b)] V_1(x,\t) }
$$-\int dz   {{\e^{\a\b\c\d}}\over 6}
 p_\a ~\N_\b\N_\c\N_\d  V_2(x,\t)$$
where  $\N_\a = d/d\t^\a$ and $\p^{\a\b}=\s_m^{\a\b}~ d/dx_m$.
Note that the only $\rho+i\s$
dependence in the integrated vertex operator of \udef\
is the $e^{-\rho-i\s}$ factor in the first term.
All terms proportional to 
$e^{n(\rho+i\s)}$ for $n>0$ must vanish on-shell in order that
$U$ has no poles with $G^-$.

The closed massless compactification-independent
vertex operator is obtained by taking
the ``square'' of the open vertex operator, i.e. 
\eqn\intver{U= \int dz d\bar z 
\bar G^- G^- \bar G^+ G^+ V(x,\t,\tb,\rho+i\s, 
\bar\rho+i\bar\s)}
where $V= \sum_{n,\bar n}
e^{n(\rho+i\s) +\bar n(\bar\rho+i
\bar\s)}
 V_{n,\bar n}(x,\t,\bar\t)$. 

\newsec{Hybrid formalism with harmonic constraints}

\subsec{Action and N=2 constraints}

Since the vertex operator of \intver\ depends only on four $\t$'s and
four $\bar\t$'s, it is not obvious how to relate $V$ and $U$ of 
\intver\ with
the standard six-dimensional Type II superfields which depend
on eight $\t$'s and eight $\bar\t$'s. 
As will be explained below, this relation will become obvious after
introducing worldsheet fields for four new $\t$'s and four new $\tb$'s
(and their conjugate momenta), as well as introducing ``harmonic''
constraints which allow half of the $\t$'s and $\tb$'s
to be gauged away. Unlike
the other worldsheet variables and the N=2 superconformal constraints,
these
new fermionic variables and harmonic
constraints do not seem to come from
worldsheet variables or constraints in the RNS formalism.

The starting point will be a free action containing the new fermionic
worldsheet variables in addition to the variables of \freeaction:
\eqn\secondaction{S=\int d^2 z (\half
 \p x^m \bar\p x_m + p_{\a j}\bar\p\t^{\a j} + \bar p_{\bar \a j}
\p\bar \t^{\bar \a j} ) +S_B+ S_C}
where $j=1$ to 2 so one has twice as many $\t$'s and $p$'s. 
The OPE's of the free fields in this action are:
\eqn\OPfree{ x^m(y) x^n(z) \to \eta^{mn} log(y-z),\quad
\rho(y) \rho(z) \to - log(y-z),\quad
\s(y) \s(z) \to - log(y-z),}
$$p_{\a j}(y) \t^{\b k}(z)\to \d_\a^\b \d_j^k (y-z)^{-1}.$$
The action of \secondaction\ can be written
in manifestly
N=2 D=6 supersymmetric notation as 
\eqn\susyaction{S=\int d^2 z (\half
 \Pi_z^m \Pi_{\bar z m} + B_{MN}^{flat} \p y^M \bar\p y^N 
+ d_{\a j}\bar\p\t^{\a j} + \bar d_{\bar \a j}
\p\bar \t^{\bar \a j})   +S_B+ S_C}
where $\Pi^m = \p x^m - {i\over 2}\e_{jk} (\s^m_{\a\b}\t^{\a j} \p\t^{\b k}
+\s^m_{\bar \a\bar \b}\tb^{\bar \a j} \p\tb^{\bar \b k})$, 
$y^M = (x^m,\t^{\a j}, \bar\t^{\bar \a j})$,
$B_{MN}^{flat} \p y^M \bar\p y^N$ is the same Wess-Zumino term as
in the Green-Schwarz action in a flat background, and 
\eqn\defd{d_{\a j} = p_{\a j} -{i\over 2}\e_{jk} \t^{ \b k} \p x_{\a\b} +
{1\over 8}\e_{\a\b\c\d}\e_{jk}\e_{lm}\t^{\b k}\t^{\c l}\p\t^{\d m} + ... ,}
$$\bar d_{\bar \a j} = \bar p_{\bar \a j} -{i\over 2}\e_{jk}
\bar \t^{\bar \b k} \bar\p x_{\bar \a \bar \b} +
{1\over 8}\e_{\bar \a\bar \b\bar \c\bar \d}\e_{jk}\e_{lm}
\tb^{\bar \b k}
\tb^{\bar \c l}\bar\p\tb^{\bar \d m} + 
 ...$$
where $...$ signifies terms which vanish using the equations of motion
(e.g. terms involving $\bar \p\t^{\a j}$ or $\p\tb^{\bar \a j}$).
Note that the non-linear terms in $d_{\a j}$ and $\bar d_{\bar \a j}$
cancel the cubic and quartic terms in \susyaction\ coming from
$\Pi_z^m \Pi_{\bar z m}$ and from the Wess-Zumino term.
In ten dimensions, a similar action to \susyaction\ was constructed by
Siegel in \ref\clas{W. Siegel,
{\it Classical Superstring Mechanics}, Nucl. Phys. B263 (1986) 93.}, 
and in four dimensions, a similar action
was constructed in \siegel. 
The first two terms of 
\susyaction\ is the standard six-dimensional GS action
in a flat background.

It is easy to check that the equations of motion
imply that $d_{\a j}$, $\Pi_z^m$, and $\p\t^{\a j}$
are holomorphic, commute with the spacetime-supersymmetry
generators,
and satisfy the
OPE's\clas
\eqn\OPE{d_{\a j}(y) d_{\b k}(z) \to -i(y-z)^{-1} \e_{jk} (\s_m)_{\a\b}
\Pi_z^m(y).}
$$d_{\a j}(y) \Pi^m_z(z) \to -i
(y-z)^{-1} \e_{jk}(\s^m)_{\a\b}\p\t^{\b k}(y),$$
$$d_{\a j}(y) \p\t^{\b k}(z) \to (y-z)^{-2} \d^k_j \d_\a^\b,\quad
\Pi_z^m(y) \Pi_z^n(z) \to (y-z)^{-2} \eta^{mn}.$$

To make \susyaction\ equivalent to the original action of \freeaction,
one now imposes the following eight first-class constraints:
\eqn\dcon{D_\a\equiv d_{\a 2} - e^{-\rho-i\s} d_{\a 1}=0,\quad 
\bar D_{\bar \a}\equiv \bar d_{\bar \a 2} - 
e^{-\bar\rho-i\bar\s}
\bar d_{\bar \a 1}=0.}
It is interesting to note that
similar constraints were used in \sok\ to describe the $D=6$ superparticle
with worldline supersymmetry.
Since $\{D_\a,\t^{\b 2}\}=\d_\a^\b$,
the first-class
constraints of \dcon\ can be used to gauge-fix $\theta^{\a 2} =
\bar \t^{\bar \a 2}=0$.
In this gauge, the action of \secondaction\ reduces to the action
of \freeaction\ where $\t^{\a 1}$ is identified with $\t^\a$ and
$p_{\a 1}$ is identified with $p_\a$. 

The N=2 superconformal generators for \secondaction\ are 
modified from those 
of 
\freeconstraints\ to:
$$T=   
{1\over 2} \Pi_z^{m} \Pi_{z~m} +
d_{\a 1}\p \t^{\a 1} +
e^{-\rho-i\s}d_{\a 1}\p \t^{\a 2} +
\half\p\rho\p\rho
+\half\p\sigma\p\sigma
+{3\over 2}\p^2 (\rho+i\sigma)+T_{C},$$
\eqn\seconstraints{ G^{+} = 
- e^{-2\rho-i\sigma} (d_1)^4 ~ +{i\over 2}
e^{-\rho} (d_{\a 1} d_{\b 1} \Pi_z^{\a\b} -2i \p(\rho+i\s) d_{\a 1}\p\t^{\a
2} +i d_{\a 1}\p^2 \t^{\a 2})}
$$
+ e^{i\sigma}( \half\Pi_z^m \Pi_{z~m} +
d_{\a 1}\p \t^{\a 1}
  +\half\p(\rho+i\sigma)\p(\rho+i\sigma)-\half\p^2(\rho+i\sigma))
+ G^{+}_{C} , $$
$$G^{-}  =
e^{-i\sigma}+ G^{-}_C ,$$
$$J=\p(\rho+i\sigma)~+J_{C} $$
where $(d_1)^4={1\over 24}\e^{\a\b\c\d}d_{\a 1} d_{\b 1} d_{\c 1} d_{\d 1}$.
When $\theta^{\a 2} =\bar \t^{\bar \a 2}=0$ and
$[\t^{\a 1},p_{\a 1}]$ are identified with $[\t^\a,
p_{\a}]$, these constraints reduce to those of \freeconstraints. 
So \freeconstraints\ can be interpreted as a gauge-fixed version
of \seconstraints. 

Note that 
${1\over 2}\Pi_z^{m} \Pi_{z~m} +
d_{\a 1}\p \t^{\a 1} +
e^{-\rho-i\s}d_{\a 1}\p \t^{\a 2}$ 
$=$
${1\over 2} \p x^m\p x_m  +
p_{\a j}\p \t^{\a j} -
D_{\a}\p \t^{\a 2}$, so $T$ is the expected free
stress-tensor for \secondaction\ when $D_\a=0$. 
The modifications of \seconstraints\ have been chosen
such that 
the N=2
constraints have no singularities
with the harmonic constraints
of \dcon. 
The absence of singularities with $G^+$ is quite amazing 
and comes from the fact that 
$G^+$ can be written as  
\eqn\gamaz{G^+ = -{1\over 24}
\epsilon^{\a\b\c\d} D_\a (D_\b (D_\c (D_\d (~e^{2\rho+3i\s}~)))) ~+G_C^+}
where $D_\a(Y)$ denotes the contour integral of $D_\a$ around $Y$.
Furthermore, the generators of \seconstraints\ still form a $c=6$ N=2 
superconformal algebra. This algebra is guaranteed since the constraints
of \seconstraints\
are invariant under the gauge transformations generated by \dcon\
and in the gauge $\t^{\a 2}=\tb^{\bar \a 2}=0$, they
reduce to the N=2 constraints of \freeconstraints.
So the free action of \secondaction, together with the constraints
of \dcon\ and \seconstraints, still describes a
critical N=2 superconformal field
theory.

\subsec{Massless compactification-independent vertex operators}

To see how the harmonic constraints of \dcon\ affect
the massless open string 
vertex operator of \intopenver, consider a function
$V$ which depends of the zero modes of
$(x^m, \t^{\a j},\rho+i\s)$
and which satisfies 
\eqn\harmcon{(\N_{\a 2}
 - e^{-\rho-i\s} \N_{\a 1})V=0, }
i.e. $V$ has no poles with \dcon\ where
$\N_{\a j}={\p\over{\p\t^{\a j}}} -{i\over 2} \e_{jk}\t^{\b k} \p_{\a\b}$. 
Defining 
\eqn\abc{\hat x^m=x^m +{i\over 4}
\s^m_{\a\b}(  e^{\rho+i\s}\t^{\a 1}\t^{\b 1}
- 
e^{-\rho-i\s}\t^{\a 2}\t^{\b 2}), }
$$\t^{\a -} = \t^{\a 1} - e^{-\rho-i\s}\t^{\a 2},\quad
\t^{\a +} = \t^{\a 1} + e^{-\rho-i\s}\t^{\a 2},$$
\harmcon\ implies that $V$ is a function of 
the zero modes of 
$(\hat x^m, \t^{\a + }, \rho+i\s)$,
but is independent of the zero modes of $\t^{\a -}$.
So the component fields of  $V$ can be related to the component
fields of $V$ in \intopenver\ by identifying $\hat x$ with $x$ and
$\t^{\a +}$ with $\t^\a$.

Applying
$G^- G^+$ on $V$
to obtain the integrated version
of the vertex operator, one gets
\eqn\secudef{ U = \int dz 
 [-  {{\e^{\a\b\c\d}}\over 6}
e^{-\rho-i\s} d_{\a 1} ~(\N_{\b 1}\N_{\c 1}\N_{\d 1} ) +i
  d_{\a 1} ~(\N_{\b 1} \partial^{\a\b} ) +{i\over 2}
 \Pi_z^{\a\b} ~(\N_{\a 1}\N_{\b 1}) }
$$+ \p\t^{\a 2}~\N_{\a 1}]V_1(x,\t)~
-\int dz   {{\e^{\a\b\c\d}}\over 6}
 d_{\a 1} ~\N_{\b 1}\N_{\c 1}\N_{\d 1} 
 V_2(x,\t).$$
Using 
\harmcon\ to relate $\N_{\a 1}V_1=\N_{\a 2}V_0$ and
$\N_{\a 1}V_2=\N_{\a 2}V_1$, $U$ can be written in a more
symmetric form in terms of $V_0$ as 
\eqn\thrudef{ U = \int dz 
 [-  {{\e^{\a\b\c\d}}\over 6}
(e^{-\rho-i\s} d_{\a 1} ~(\N_{\b 1}\N_{\c 1}\N_{\d 2} ) 
+  d_{\a 1} ~(\N_{\b 2}\N_{\c 2}\N_{\d 1} )) }
$$
  +{i\over 4}
 \Pi_z^{\a\b} ~[\N_{\a 1},\N_{\b 2}] -\half
\p\t^{\a 1}~\N_{\a 1}+\half
\p\t^{\a 2}~\N_{\a 2}]V_0(x,\t)
$$
where we have subtracted the surface term
$\half\int dz [\Pi_z^m\p_m+ \p\t^{\a j}\N_{\a j}]V_0 =$
$\half\int dz \p V_0$.
Replacing $e^{-\rho-i\s} d_{\a 1}$ with $d_{\a 2}$ using
\dcon, the  
vertex operator of \thrudef\ 
closely resembles the
four-dimensional massless vertex operator of \ref\four
{N. Berkovits, {\it Covariant Quantization of the Superstring
in a Calabi-Yau Background}, Nucl. Phys. B431 (1994) 258,
hep-th/9404162.}. It is interesting to note that both the
four and six-dimensional massless 
vertex operators are of the form
proposed by Siegel in \clas\ for the ten-dimensional vertex operator,
$U=\int dz [d_\a W^\a + \p y^M A_M]$
where $W^\a$ is the super-Yang-Mills field-strength and 
$A_M$ are the superspace gauge fields.

The closed massless vertex operator is the ``square'' of \thrudef, i.e.
\eqn\thrucl{ U = \int d^2 z 
 |
 -  {{\e^{\a\b\c\d}}\over 6}
(d_{\a 2} ~(\N_{\b 1}\N_{\c 1}\N_{\d 2} ) 
+  d_{\a 1} ~(\N_{ \b 2}\N_{\c 2}\N_{\d 1} )) }
$$
  +{i\over 4}
 \Pi_z^{\a\b} ~[\N_{\a 1},\N_{\b 2}] -
\half\p\t^{\a 1}~\N_{\a 1}+\half
\p\t^{\a 2}~\N_{\a 2}
 |^2 ~
V_{0,0}(x,\t,\tb). $$

\newsec{Hybrid action in a curved background}

Given the N=2 D=6 spacetime-supersymmetric form of the action
in a flat background in \susyaction\ and the closed
massless vertex operators of \thrucl, it is easy to guess
the action in a curved background if one ignores the Fradkin-Tseytlin
term which couples the dilaton to the worldsheet curvature. 
This action is given by
\eqn\curved{
S_0={1\over {\a '}}\int d^2 z (\half
 \Pi_z^c \Pi_{\bar z c} 
+ B_{MN} \p y^M \bar\p y^N + 
 d_{\a j}\Pi_{\bar z}^{\a j}
+ \bar d_{\bar \a j} \Pi_z^{\bar \a j}
+ d_{\a j} \bar d_{\bar\b k} P^{\a j~\bar \b k} )
+ S_B+ S_C}
where $E_M^A$ is the super-vierbein with $A=(c,\a j,\bar\a j)$
and $M=(m,\mu j,\bar\mu j)$, $\Pi^c = E_M^c \p y^M$ is the
vector current, $\Pi^{\a j }= E_M^{\a j} \p y^M$
and  
$\Pi^{\bar\a j} = E_M^{\bar\a j} \p y^M$ are the spinor currents,
and $P^{\a j~\bar\b k}$ is the superfield whose
lowest components are the bispinor Ramond-Ramond field strengths.
The necessity of including the term proportional to $d_{\a j}
\bar d_{\bar \a k}$ can be seen from the massless vertex operator of 
\thrucl.

To reproduce the expected coupling-constant dependence
of scattering amplitudes, one needs a Fradkin-Tseytlin term
which couples the spacetime dilaton to the worldsheet curvature. 
As in the four-dimensional sigma model of \siegel, this term
is constructed by coupling spacetime compensator 
superfields to the N=(2,2) worldsheet supercurvature. 
The N=(2,2) worldsheet supercurvature is described by
a chiral and twisted-chiral worldsheet superfield $\Sigma_c$ and
$\Sigma_{tc}$, and their complex conjugates $\Sigma_c^*$
and $\Sigma_{tc}^*$. 
In the four-dimensional sigma model, the worldsheet supercurvature coupled
to spacetime superfields $\Phi_c$ and $\Phi_{tc}$ 
which compensated a $U(1)\times U(1)$ subgroup of the
$U(1)\times SU(2)$ R-transformations of 
N=2 D=4 supergravity. In the six-dimensional sigma model,
the worldsheet supercurvature superfields couple to 
spacetime superfields $\Phi_c$ and $\Phi_{tc}$ 
which compensate a $U(1)\times U(1)$ subgroup of the
$SU(2)\times SU(2)$ R-transformations of 
N=2 D=6 supergravity. 
In both the four and
six-dimensional actions, the coupling term is defined as
\eqn\ft{S_{FT}= \int d^2 z [G^- \bar G^- (\Phi_c \Sigma_c) 
+ G^- \bar G^+ (\Phi_{tc} \Sigma_{tc}) 
+G^+ \bar G^+ (\Phi_c^* \Sigma_c^*) 
+ G^+ \bar G^- (\Phi_{tc}^* \Sigma_{tc}^*)]} 
where $\Phi^*_c$ and $\Phi^*_{tc}$ are the complex conjugates of
$\Phi_c$ and $\Phi_{tc}$.
The complete sigma model action is therefore $S=S_0+S_{FT}$.

The spacetime compensator superfields $\Phi_c$ and $\Phi_{tc}$
are functions of the
zero modes of $(x^m,\t^{\mu j},$
$\bar\t^{\bar \mu j}, \rho+i\s,
\rho+i\bar\s)$ which satisfy the chirality and twisted-chirality
constraints
\eqn\chcon{G^+\Phi_c =\bar G^+\Phi_c=0,\quad
G^-\Phi_c^* =\bar G^-\Phi_c^*=0,}
$$G^+\Phi_{tc} =\bar G^-\Phi_{tc}=0,\quad
G^-\Phi_{tc}^* =\bar G^+\Phi_{tc}^*=0,$$
in addition to the constraints implied by \dcon\ that
\eqn\dimpl{(\N_{\a 1} - e^{-\rho-i\s}\N_{\a 2})\Phi=
(\N_{\a 1} - e^{-\rho-i\s}\N_{\a 2})\Phi^*=0,}
$$
(\bar\N_{\bar\a 1} - e^{-\bar\rho-i\bar\s}
\bar\N_{\bar\a 2})\Phi=
(\bar\N_{\bar\a 1} - e^{-\bar\rho-i\bar\s}
\bar\N_{\bar\a 2})\Phi^*=
0.$$
Writing $\Phi=\Sigma_{n,\bar n}  e^{n(\rho+i\s) +
\bar n(\bar\rho+i\bar\s)} \Phi^{n,\bar n}$, \chcon\ and \dimpl\ imply that
$\Phi_c^{n,\bar n}=0$ when either $n<0$ or $\bar n<0 $ and
$\Phi_{tc}^{n,\bar n}=0$ when either $n<0$ or $\bar n>0$.
To see that $G^+ \Phi_c=0$ implies that 
$\Phi_c^{n,\bar n}=0$ when $n<0$, observe that \gamaz\ and
\dimpl\ imply that $\Phi_c$ has no poles with $e^{2\rho+3i\s}$.
The complex conjugate superfields $\Phi_c^*$ and $\Phi_{tc}^*$ are
related to 
$\Phi_c$ and $\Phi_{tc}$ by defining
$(\t^{\a j})^*=\e^{jk}\t^{\a k}$,
$(\tb^{\bar\a j})^*=\e^{jk}\tb^{\bar a k}$,
$(e^{\rho+i\s})^* = - e^{-\rho-i\s}$ and
$(e^{\bar\rho+i\bar\s})^* = - e^{-\bar\rho-i\bar\s}$. One can
check that this definition of complex conjugation implies
that $\Phi_c^*$ and $\Phi_{tc}^*$ satisfy \chcon\ and \dimpl.

The N=2 constraints for the superstring in a curved six-dimensional
background
are given by 
$$T= T_0 +  \p^2 (\Phi_c +\Phi_c^* +
\Phi_{tc} +\Phi_{tc}^* )
,$$
\eqn\concurve{ G^{+} = G^+_0 + G^+ \p(\Phi_c^* +\Phi_{tc}^*),\quad
 G^{-} = G^-_0 + G^- \p(\Phi_c +\Phi_{tc}),}
$$J=J_0 +\p(
\Phi_c -\Phi_c^* +
\Phi_{tc} -\Phi_{tc}^* ),$$
where $[T_0,G^+_0,G^-_0,J_0]$ are the N=2 constraints
of \seconstraints\ after replacing $\Pi_z^m$ with $\Pi_z^c$
and $\p\t^{\a j}$ with $\Pi^{\a j}$. 
The Fradkin-Tseytlin contribution to the
constraints of \concurve\ are analogous to those discussed in
\siegel\ and \ref\deboer{J. de Boer and K. Skenderis,
{\it Covariant Computation of the Low Energy Effective Action of
the Heterotic Superstring}, Nucl. Phys. B481 (1996) 129, hep-th/9608078.} for
the four-dimensional background.

\newsec{ Relation to $PSU(2|2)$ action for $AdS_3\times S^3$}

In reference \VW, an action based on the $PSU(2|2)$ supergroup
was constructed for the superstring in an $AdS_3\times S^3$
background with Ramond-Ramond flux. Although the $(z,\bar z)$
symmetric part of this action was the usual group action
for $PSU(2|2)$, the $(z,\bar z)$ anti-symmetric part of this
action contained complicated dependence on the chiral and
anti-chiral bosons $\rho+i\s$ and $\bar \rho+i\bar\s$. The N=2
worldsheet superconformal generators also depended in
a complicated way on these bosons. When the paper was
written, it was unclear how to give a geometrical interpretation
to this $(\rho+i\s,\bar\rho+i\bar\s)$ dependence. 

Using the results of the previous section, it will now
be shown how the rather complicated $(\rho+i\s,\bar\rho+i\bar\s)$ dependence
in the action of \VW\ can be derived from a relatively
simple action whose only dependence on $\rho+i\s$ and $\bar\rho+i\bar\s$
comes from the harmonic constraints of \dcon.

Since the $AdS_3\times S^3$ background
of interest has a constant dilaton, the Fradkin-Tseytlin
term of \ft\ contributes the usual Euler number dependence which
will be ignored. The remaining part of the action is given
by \curved\
where $E^A_M$, $B_{MN}$ and $P^{\a j~\bar\b k}$ take values
determined by the $AdS_3\times S^3$ metric and by the NS-NS and R-R 
three-form flux.

For convenience, only backgrounds with
pure Ramond-Ramond flux will be discussed
although it should be easy to generalize the discussion to include
backgrounds with NS-NS flux.
In the presence of a three-form R-R flux with values\foot
{This choice
of three-form R-R flux is slightly more convenient that that of \VW\ since
it preserves a diagonal SU(2) $R$-symmetry.} 
\eqn\hdef{H^{012}_{jk}=H^{345}_{jk}=N\e_{jk},}
the superfield 
$P^{\a j~\bar\b k}$ satisfies
\eqn\defpp{
P^{\a j~\bar\b k}=
N\l \d^{\a\bar\b} \e^{jk}}
where the coupling constant
$\l$ appears in \defpp\ because there is no $\l^{-2}$ factor in front of
the Ramond-Ramond $H^{mnp}H_{mnp}$ kinetic term in the action.
Furthermore, in the $AdS_3\times S^3$
background, one can choose the only non-zero values of $B_{AB}=
E_A^M E_B^N B_{MN}$  to be \BBZ
\eqn\defB{ B_{\a j~\bar\b k}= B_{\bar\b k~\a j} = 
-{1\over 4} (N\l)^{-1} \e_{jk} \d_{\a\bar\b}.}
Using $H_{ABC} = \nabla_{[A} B_{BC]} + T_{[AB}{}^D B_{C]D}$
and the torsion constraints
$$T_{c ~\a j}{}^{\bar\c k} = (\s_c)_{\a\b} \e_{jl}P^{\b l~\bar\c k},\quad
T_{c ~\bar\a j}{}^{\c k} = -(\s_c)_{\bar\a\bar\b} \e_{jl}P^{\c k~\bar\b l},$$
it is easy to show that 
$H_{c~\a j~ \b k}= {1\over 4}(\s_c)_{\a\b} \e_{jk}$ 
and
$H_{c~\bar\a j~ \bar\b k}= -{1\over 4}(\s_c)_{\bar\a\bar\b} \e_{jk}$ 
as desired.

Plugging these values for the background fields into \curved, 
one obtains
\eqn\addd{ 
S={1\over {\a '}}\int d^2 z [\half
 \Pi_z^c \Pi_{\bar z c} 
-{1\over 4}
(N\l)^{-1} \d_{jk}\d_{\a\bar \b}
(\Pi_z^{\a j}\Pi_{\bar z}^{\bar \b k}-
\Pi_{\bar z}^{\a j}\Pi_{z}^{\bar \b k})}
$$+ d_{\a j}\Pi_{\bar z}^{\a j}
+ \bar d_{\bar \a j} \Pi_z^{\bar \a j} 
+ \e^{jk} \d^{\a\bar\b} N\l d_{\a j} \bar d_{\bar\b k}]
+S_B +S_C.$$
After rescaling $E_M^c \to (N\l)^{-1} E_M^c$, 
$E_M^{\a j} \to (N\l)^{-\half} E_M^{\a j}$, 
$E_M^{\bar\a j} \to (N\l)^{-\half} E_M^{\bar\a j}$, 
$d_{\a j} \to (N\l)^{-{3\over 2}}d_{\a j}$, 
$\bar d_{\bar \a j} \to (N\l)^{-{3\over 2}}\bar d_{\bar \a j}$, 
the $(N\l)$ dependence of \addd\ simplifies to 
\eqn\aee{ 
S={1\over {\a 'N^2\l^2}}\int d^2 z [\half
 \Pi_z^c \Pi_{\bar z c} -{1\over 4}
\e_{jk}\d_{\a\bar \b}
(\Pi_z^{\a j}\Pi_{\bar z}^{\bar \b k}-
\Pi_{\bar z}^{\a j}\Pi_{z}^{\bar \b k})}
$$+
d_{\a j}\Pi_{\bar z}^{\a j}
+ \bar d_{\bar \a j} \Pi_z^{\bar \a j} + 
\e^{jk} \d^{\a\bar\b} d_{\a j} \bar d_{\bar\b k}]
+  S_B+S_C.$$
The first line of \aee\ is precisely
the GS action for the 
$AdS_3\times S^3$ background which was studied in \ref\tsthree
{J. Rahmfeld and A. Rajaraman, {\it The GS String Action On ${\rm AdS}_3\times
S^3$
with Ramond-Ramond Charge}, hep-th/9809164 \semi
J. Park and S.-J. Rey, {\it
Green-Schwarz Superstring on ${\rm AdS}_3\times
{S}^3$}, hep-th/9812062\semi
I. Pesando, {\it The GS Type IIB Superstring Action on ${\rm AdS}_3\times S^3
\times {T}^4$}, hep-th/9809145.}
and which is based on the $AdS_5\times S^5$ action of \ref\ts{
R. Metsaev and
A. Tseytlin, 
{\it Type IIB superstring action in $AdS_5 \times S^5$ 
background}, 
Nucl.Phys. {B533} (1998) 109, hep-th/9805028.}.
However, the second line of \aee\ is crucial for quantization
and is absent from the action of \tsthree.

As discussed in \tsthree\ts, the currents $[\Pi^c,\Pi^{\a j},
\Pi^{\bar\a j}$] can be identified with the 
currents of the
supergroup coset 
 $PSU(2|2)\times PSU(2|2)/ SU(2)\times SU(2)$. If $g$ takes
values in the supergroup
$PSU(2|2)\times PSU(2|2)$, the left-invariant one-forms $g^{-1} \p g$
can be defined as $(S^{\a j},K^{\a\b})$ and $(\tilde S^{\a j},
\tilde K^{\a\b})$ which generate the Lie algebra
\eqn\intu{\eqalign{ [K^{\a\b},K^{\c\d}] & = 
\delta^{\a\c}K^{\b\d}-\delta^{\a\d}K^{\b\c}
              -\delta^{\b\c}K^{\a\d}+\delta^{\b\d}K^{\a\c}  \cr
                    [K^{\a\b},S^{\c j}] & = \delta^{\a\c}S^{\b j}
                                   -\delta^{\b\c}S^{\a j}\cr
                    \{S^{\a j},S^{\b k}\}& = \half \epsilon^{jk}
                         \epsilon^{\a\b\c\d}K^{\c\d},\cr}}
and similarly for the tilded currents. (The untilded and tilded
currents commute with each other.) Defining $K^{\a\b}+\tilde K^{\a\b}$
to be the one-forms which are absent from the action to provide the
local $SU(2)\times SU(2)$ invariance, the remaining six bosonic
currents and sixteen fermionic currents are related to
$[\Pi^c,\Pi^{\a j},
\Pi^{\bar\a j}$] as 
\eqn\currel{\Pi^c =\s^c_{\a\b} (K^{\a\b} -\tilde K^{\a\b}),\quad
\Pi^{\a j}= S^{\a j} +i\tilde S^{\a j},\quad
\Pi^{\bar\a j}= S^{\a j} -i\tilde S^{\a j}}
where $(x,\t,\tb)$
parameterize the 
$PSU(2|2)\times PSU(2|2)/ SU(2)\times SU(2)$
coset supermanifold.

To further simplify the action of \aee, one has two options. 
One option is to integrate out all
$d_{\a j}$ and $\bar d_{\bar \a j}$ worldsheet fields, producing
the action 
\eqn\optone{S={1\over {\a ' N^2\l^2}}\int d^2 z [\half
 \Pi_z^c \Pi_{\bar z c} - 
\e_{jk}\d_{\a\bar \b}
({1\over 4}\Pi_z^{\a j}\Pi_{\bar z}^{\bar \b k}+{3\over 4}
\Pi_{\bar z}^{\a j}\Pi_{z}^{\bar \b k}) ] +S_B+ S_C.}
Using the identification of \currel, the above action
can be interpreted as a sigma model action with WZ term
for the supercoset
$PSU(2|2)\times PSU(2|2)/ SU(2)\times SU(2)$. 
In fact, precisely this sigma model
action was considered in \BBZ\ and shown to be one-loop
conformally invariant. Of course, this action needs
to be supplemented by the constraints of \dcon\ and
\seconstraints\  to remove the unphysical
degrees of freedom.

A second option is to first use the eight \dcon\ constraints
and the six local $SU(2)\times SU(2)$ invariances to 
gauge-fix to the identity the 
tilded $PSU(2|2)$ group parameters. 
This gauge-fixing can be inserted directly into the
action of \aee\ since the group parameters transform without
derivatives under these gauge transformations.
The resulting action (after imposing the \dcon\ constraints
to solve for $d_{\a 2}$ and $\bar d_{\bar\a 1}$ in terms
of $d_{\a 1}$ and $\bar d_{\bar \a 2}$) is
\eqn\agg{ 
S={1\over {\a 'N^2\l^2}}\int d^2 z [{1\over 8}
\e_{\a\b\c\d} K_z^{\a\b} K_{\bar z}^{\c\d} }
$$+
d_{\a 1}(S_{\bar z}^{\a 1} + e^{-\rho-i\s}S_{\bar z}^{\a 2})
+ \bar d_{\a 2} (S_{z}^{\a 2} 
+ e^{\bar\rho+i\bar\s}S_{z}^{\a 1})
+
(1- e^{-\rho-i\s+\bar\rho+i\bar\s})  d_{\a 1} \bar d_{\a 2}]
+  S_B+ S_C.$$

Integrating out $d_{\a 1}$ and $\bar d_{\a 2}$, one obtains
\eqn\ahh{ 
S={1\over {\a 'N^2\l^2}}\int d^2 z [{1\over 8}
\e_{\a\b\c\d} K_z^{\a\b} K_{\bar z}^{\c\d}  }
$$
-(1- x)^{-1}
(S_{\bar z}^{\a 1} + e^{-\rho-i\s}S_{\bar z}^{\a 2})
(S_{z}^{\a 2} 
+ e^{\bar\rho+i\bar\s}S_{z}^{\a 1})]
+  S_B + S_C$$
\eqn\ajj{ 
={1\over {\a 'N^2\l^2}}\int d^2 z [{1\over 8}
\e_{\a\b\c\d} K_z^{\a\b} K_{\bar z}^{\c\d} -\half
\e_{jk} S_z^{\a j} S_{\bar  z}^{\a k} }
$$
+(1-x)^{-1} 
( 
 e^{\bar\rho+i\bar\s}
S^{\a 1}\wedge S^{\a 1}
 + e^{-\rho-i\s}S^{\a 2} \wedge S^{\a 2} 
 +2
S_z^{\a 1}\wedge S_{\bar z}^{\a 2}) ] +S_B +S_C,$$
where
$x=
e^{-\rho-i\s+\bar\rho+i\bar\s}$ and a term proportional to
$\int d^2 z ~S^{\a 1}\wedge S^{\a 2}$ (with no
$\rho+i\s$ factors) has been dropped since it is
a total derivative \VW.

This action is the same as that of \VW\ if one redefines
$\rho\to\rho +log(2)$,
$\bar\rho\to -\bar\rho -log(2)$, $\s \to \s$ and $\bar\s \to -\bar\s$.
The N=2 constraints of \concurve\ can be related to those of \VW\
by inserting the equation of motion for $d_1$ into \concurve\
in the gauge where all tilded currents vanish. The result is
$$T=   
{1\over 8} \e_{\a\b\c\d}K^{\a\b} K^{\c\d}-\half \e_{jk}S^{\a j}S^{\a k}
+\half\p\rho\p\rho
+\half\p\sigma\p\sigma
+{3\over 2}\p^2 (\rho+i\sigma)+T_{C},$$
\eqn\concurve{ G^{+} = 
- e^{-2\rho-i\sigma} (1-x)^{-4} (S^2+e^{\bar\rho+i\bar\s}S^1)^4 ~ }
$$+{i\over 2}
e^{-\rho} [(1-x)^{-2}(
S^{\a 2}+e^{\bar\rho+i\bar\s}S^{\a 1})
(S^{\b 2}+e^{\bar\rho+i\bar\s}S^{\b 1}) K^{\a\b}$$
$$
-2i \p(\rho+i\s) 
e^{\bar\rho+i\bar\s}(1-x)^{-1}S^{\a 1}
S^{\a 2}+ i(1-x)^{-1}
(S^{\a 2}+e^{\bar\rho+i\bar\s}S^{\a 1})\p S^{\a 2}]
$$
$$
+e^{i\sigma}( 
{1\over 8} \e_{\a\b\c\d}K^{\a\b} K^{\c\d}+(1-x)^{-1}S^{\a 2}S^{\a 1}
  +\half\p(\rho+i\sigma)\p(\rho+i\sigma)-\half\p^2(\rho+i\sigma))
+ G^{+}_{C} , $$
$$G^{-}  =
e^{-i\sigma}+ G^{-}_C ,$$
$$J=\p(\rho+i\sigma)~+J_{C}, $$
which matches the N=2 constraints of $\VW$ up to normal-ordering
ambiguities (after redefining
$\rho\to\rho +log(2)$,
$\bar\rho\to -\bar\rho -log(2)$, $\s \to \s$ and $\bar\s \to -\bar\s$).

{\bf Acknowledgements:} I would like to thank Michael
Bershadsky, Louise Dolan, Paul Howe, Jaemo Park,
Warren Siegel, Emery Sokatchev, Cumrun Vafa and Edward
Witten for useful discussions. I would also like
to thank the IAS at Princeton,
Harvard University, the ICTP at Trieste, the University
of Amsterdam, and Potsdam University
for their hospitality and CNPq grant 300256/94-9
for partial financial support.

\listrefs

\end